\documentclass[runningheads]{llncs}
\usepackage{graphicx}
\usepackage{cite}
\usepackage{amsmath,amssymb,amsfonts}
\usepackage{algorithmic}
\usepackage{graphicx}
\usepackage{textcomp}
\usepackage{xcolor}
\usepackage{hyperref}
\usepackage{refstyle}
\usepackage{wasysym}
\usepackage{latexsym}
\usepackage{listings}
\usepackage{float}
\usepackage{color}
\usepackage{xcolor}
\usepackage{xspace}
\usepackage{multirow}
\usepackage{rotating}
\usepackage{tcolorbox}
\newcommand\tabrotate[1]{\rotatebox{90}{\verschiebung#1}}
\newcommand\verschiebung[1][-.5\normalbaselineskip]{\hspace{#1}}
\usepackage{thmtools}

\declaretheorem[thmbox=L, name=Research Question]{rq}
\begin{document}
\title{Identifying the Mood of a Software Development Team by Analyzing Text-Based Communication in Chats with Machine Learning}
\titlerunning{Identifying the Mood of a Software Development Team}
\author{Jil Kl\"under\orcidID{0000-0001-7674-2930} \and Julian Horstmann
\and \\Oliver Karras\orcidID{0000-0001-5336-6899} }

\authorrunning{J. Kl\"under et al.}
\institute{Leibniz University Hannover, Software Engineering Group, Hannover, Germany 
\email{\{jil.kluender | oliver.karras\}@inf.uni-hannover.de, julian.horstmann@se.uni-hannover.de }}

\maketitle              % typeset the header of the contribution
\begin{abstract}
Software development encompasses many collaborative tasks in which usually several persons are involved. Close collaboration and the synchronization of different members of the development team require effective communication. One established communication channel are meetings which are, however, often not as effective as expected. Several approaches already focused on the analysis of meetings to determine the reasons for inefficiency and dissatisfying meeting outcomes. In addition to meetings, text-based communication channels such as chats and e-mails are frequently used in development teams. Communication via these channels requires a similar appropriate behavior as in meetings to achieve a satisfying and expedient collaboration. However, these channels have not yet been extensively examined in research. 

In this paper, we present an approach for analyzing interpersonal behavior in text-based communication concerning the conversational tone, the familiarity of sender and receiver, the sender's emotionality, and the appropriateness of the used language. We evaluate our approach in an industrial case study based on 1947 messages sent in a group chat in Zulip over 5.5 months. Using our approach, it was possible to automatically classify written sentences as positive, neutral, or negative with an average accuracy of 62.97\% compared to human ratings. Despite this coarse-grained classification, it is possible to gain an overall picture of the adequacy of the textual communication and tendencies in the group mood.

\end{abstract}

\section{Introduction}
Due to the increasing complexity of software, most software projects require some kind of teamwork \cite{kraut1995coordination}. Having a team working on a project requires coordination and an adequate collaboration \cite{kraut1995coordination,ghosh2004using}, for example, appropriate requirements communication between the development team and the customer. To succeed with the project, the team and the customer must share the same vision \cite{fricker2015requirements}. Otherwise, the team cannot develop a software satisfying the customer \cite{bjarnason2011requirements}. 

An adequate collaboration requires knowledge and information sharing to have a successful project closure \cite{marjaie2011communication,prenner2018making}. Lost or insufficiently shared information can cause -- in the worst case -- project failure, e.g., due to missing functionality of the final software product. Mitigating this risk requires a sufficient amount of communication during the whole development process \cite{marjaie2011communication}. This communication can take place, e.g., in meetings, via e-mail, or during phone calls \cite{kluender2017team}.

As meetings enable team members to share a lot of information with many team members in a short time, they are an established medium in the development process \cite{oshri2007global,schneider2018positive}. However, inappropriate behavior and interactions in meetings decrease the success of a meeting and the participants' satisfaction afterwards \cite{prenner2018making,schneider2018positive}. This, in turn, has an influence on the project and the collaboration. To avoid inappropriate behavior in meetings, interaction analyses are an established medium in psychology \cite{kauffeld2012meetings,lehmann2011verbal} and gain increasing attention in software engineering \cite{prenner2018making,schneider2018positive,kluender2020do}. 

In addition to the increasing complexity of software projects, the share of globally distributed projects is high \cite{Kuhrmann:2018aa} and complicates a close collaboration \cite{teasley2000does,zheng2002trust}. In case of regionally or globally distributed projects, it is difficult to have meetings regularly \cite{oshri2007global}. Virtual meetings are a possibility, which is, however, influenced by the requirements for technical equipment (including bandwidth) and difficulties caused by different time zones. Therefore, indirect communication using digital communication channels such as e-mails or instant messenger is widely used in software projects \cite{kluender2017team}. 

\textit{Problem Statement.} According to Schneider et al. \cite{schneider2018positive} and Kauffeld et al. \cite{kauffeld2009complaint}, a single person participating in a meeting can influence the mood of all other participants -- both positively or negatively. This, in turn, influences the developers' productivity \cite{graziotin2013happy} and has several other consequences for the project \cite{graziotin2017consequences}. Therefore, interaction analyses in meetings take into account the amount of positive, i.e., good and appropriate, as well as negative, i.e., bad and inappropriate, behavior during the meetings \cite{kluender2020do,kauffeld2012meetings}. Since the frequency and duration of meetings tend to decrease with project progress, whereas the use of other communication channels increases \cite{kluender2017team}, likely, the communication behavior in text messages can also influence team satisfaction and, thus, motivation and project progress. 

\textit{Objective.}
In this paper, \textit{we want to analyze text-based communication, for example in e-mails or chats, in development teams with respect to its emotionality to detect development phases where the group mood is rather negative}. The emotionality of text-based communication is for example affected by the language used, the frequency, the length of the messages, the formality of the communication, the use of emoticons, and the time until the receiver replies to the message. In particular, we want to answer the following research question: 

\begin{rq}
	How can text-based communication in development teams be holistically analyzed to derive information on the mood in the team?
\end{rq}

\textit{Contribution.}
We present the current state of our approach classifying written messages as \textit{positive},\textit{ negative}, or \textit{neutral} based on the sensitivity and the formality of the used words. We evaluate the approach in a case study in industry based on 1947 messages in a group chat. The results show that our tool classifies single sentences as \textit{positive}, \textit{negative}, or \textit{neutral} with an average accuracy of 62.97\%. When refining the analysis techniques it is possible to increase this number of correctly identified sentences either to shed light on the overall mood transported in messages or to analyze the mood in the development team based on the messages to allow interventions in case of a very dissatisfied team.  

\textit{Outline.}
The rest of the paper is structured as follows: In Section~\ref{sec:rw}, we present related work. Section~\ref{sec:researchdesign} summarizes the concept and the approach followed in this paper. We evaluate the approach in Section~\ref{sec:eval} and present the results of the application in industry in Section~\ref{sec:results} which we discuss in Section~\ref{sec:discussion}. We conclude the paper in Section~\ref{sec:conclusion}.

\section{Related Work}\label{sec:rw}
Analyzing communication behavior is not new in the area of Software Engineering. McChesney and Gallagher \cite{McChesney2004} analyze both communication and coordination in software projects. Herbsleb and Mockus \cite{Herbsleb2003} analyze differences in communication behavior of distributed and co-located teams. Kl\"under et al. \cite{kluender2017team} analyze team meetings, their frequency and duration over time in software projects. On a more fine-grained level, Schneider et al. \cite{schneider2018positive} analyze interactions in team meetings of development teams. All these analyses require manual effort. 

However, there are some approaches to support the analysis of communication of development teams by tools. Most existing approaches focus on meetings. Gall and Berenbach \cite{gall2006towards} present a framework to record elicitation meetings and automatically save information given by stakeholders. Shakeri et al. \cite{shakeri2018elica} also support the analysis of elicitation meetings. Their tool automatically collects knowledge that is important to understand the requirements. This approach is mainly based on written documentation and allows a content-related analysis. However, it has not yet been applied to written communication in text messages of development teams. 

Sentiment analysis, i.e., the analysis of textual language aiming at identifying the author's mood, is also not new. Jongeling et al. \cite{jongeling2015choosing} compare different tools for sentiment analysis in the Software Engineering domain and compare the tools' results to human evaluators. According to their results, different tools can produce contradictory results. Islam and Zibran \cite{islam2017leveraging} also analyze and compare the results of different tools used for sentiment analysis to understand their low accuracy. They present an improved version of one of the tools adjusted to development teams. The inaccuracy of tools can be partially explained by training the used classifiers on data sets which are not related to the Software Engineering domain and hence do not consider domain-typical language and knowledge. Lin et al. \cite{lin2018sentiment} trained an already existing tool for sentiment analysis using 40000 manually labeled sentences or words from Stack Overflow. Calefato et al. \cite{calefato2018sentiment} present Senti4SD which is a classifier adjusted to development teams. The classifier was trained using communication of developers on Stack Overflow. Jurado and Rodriguez \cite{jurado2015sentiment} analyze text in issues and tickets with sentiment analysis to extend the possibilities to investigate the development process.

In this paper, we use sentiment analysis to analyze text-based communication of developers on team level, which has, to the best of our knowledge, not been done before. 

\section{General Research Approach}\label{sec:researchdesign}
\begin{figure}
	\centering
	\includegraphics[width=1.0\linewidth]{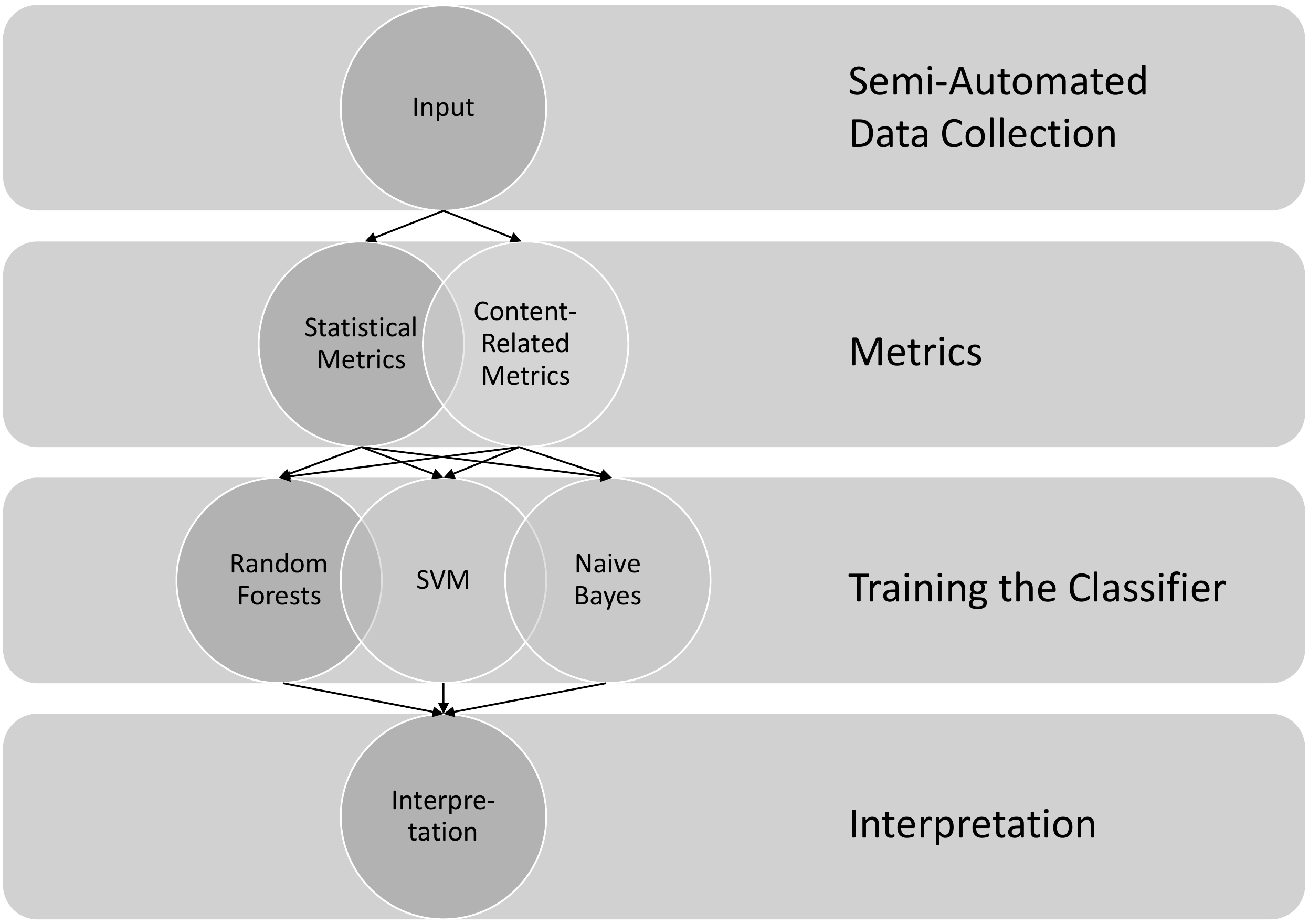}
	\caption{Overview of the process to classify messages}
	\label{fig:research-overview}
\end{figure}

To achieve our research goal, we combine sentiment analysis with natural language processing and machine learning to classify text-based communication. We developed an approach to classify text-based communication as \textit{positive, neutral} or \textit{negative} considering the emotion transported in the message. In the following, we give an overview of the process presented in Fig.~\ref{fig:research-overview}. An exemplary application of our approach is presented in Section~\ref{sec:eval}.  

Our approach consists of the four steps summarized in Fig.~\ref{fig:research-overview}. It starts with a semi-automated data collection by crawling the respective communication channel. Afterwards, different metrics are calculated for the text messages to extract relevant characteristics of the messages to allow conclusions regarding the emotionality of the communication. These metrics are used by a trained classifier to assign one of the classes \textit{positive, neutral}, or \textit{negative} to each of the sentences. This classification allows an evaluation and interpretation of the whole communication, e.g., on a daily basis. 

\subsection{Step 1: Data Collection}
Text-based communication can be found in different sources, including mailing lists, instant messengers such as Slack\footnote{\url{https://slack.com/}}, Skype\footnote{\url{https://www.skype.com/}}, or Zulip\footnote{\url{https://zulipchat.com/}}, and in e-mails. To analyze the communication on group level, it is most suitable to use some kind of group chat or mailing lists several team members have access to and use it to discuss team-internal project-related issues. However, the process is applicable to any kind of text-based communication, including bi-directional information exchange\footnote{Note that the analysis of bi-directional communication is questionable due to privacy concerns and personal messages in private chats.}.

The semi-automated data collection strongly depends on the communication channel under consideration. Exporting text messages in Skype differs from a data export in e-mails. At the moment, our approach supports the semi-automated data collection for Zulip, which will be described in more detail in Section~\ref{sec:eval} as part of the application in industry. 

\subsection{Step 2: Metrics}\label{sec:metrics}
Metrics are required to extract relevant characteristics of text-based communication. The identification of relevant metrics is difficult as natural language is not unique and can be interpreted in completely different ways \cite{watzlawick2011pragmatics}. Depending on the actual goal of the communication analysis, one may choose different metrics. In the following, we present some exemplary metrics and the rationale of why we consider them useful for emotional analysis of text-based communication.

\subsubsection{Statistical metrics} analyze the text messages from a quantitative viewpoint, for example by calculating the \textit{length of each message} or the \textit{average length of the used words}. The length of the words and the message allow conclusions for the formality of the communication. Consider the following situation: 

\begin{tcolorbox}[title=Example] 
	Paul has a problem and writes a message with 100 words in the group chat. In his text, he explains the problem in detail so that everybody who reads the message has a clear idea of what information he needs to solve the problem. In the end, he proposes a very time-consuming alternative if he does not get the required information. And the only answer he receives comes from Anton: ``sounds good''. 
\end{tcolorbox}

This huge difference in the length of the messages allows several conclusions. First, it raises the impression that nobody but Anton cares about Paul's problem. Second, as Anton's reaction is short, he may not even have read the whole text, or is at least not interested in supporting Paul. Of course, one can also interpret the short message differently. But this is one scenario, where a short message can lead to demotivation and team-internal problems. And it is not clear how Paul interprets this answer. 

Besides the lengths of messages and words, counting adjectives, emoticons and the number of punctuation characters is also useful. The use of emoticons indicates familiarity in the team. In formal communication, emoticons will only be partially used, if at all. The number of punctuation characters is difficult to interpret. On the one hand, the use of commas and the like imply a formal message, but it can also indicate a huge amount of emoticons such as ``:-)''. Nonetheless, in conjunction with a detection of smileys using punctuation characters can help to analyze the formality of the message. 

\subsubsection{Content-Related Metrics}
The statistical metrics do not allow conclusions on the emotionality of the messages which is the main topic of this paper. Therefore, the content of text-based communication has also to be taken into account. To analyze the communication on content-level, profound knowledge on a typical structure of the language, the so-called \textit{part of speech}, is necessary. Using natural language processing, it is possible to analyze words in relation to their position in the sentence. This allows considering not only the word $A$ itself but also other words that may influence the interpretation of the word $A$. 

\textit{Emoticons} also raise feelings, which may differ among the receiver of a message \cite{wang2015sentiment}. Wang and Castanon \cite{wang2015sentiment} investigated the use of emoticons and the intention of the use. According to their results, it is not always possible to assign exactly one of the classes \textit{positive, neutral}, or \textit{negative} to the emoticon \cite{wang2015sentiment}. Therefore, they present a list of emoticons together with a probability that the emoticon belongs to the respective class. For example, the emoticon ``:D'' is interpreted positively with a probability of 90\%. In the approach presented in this paper, we assign the class with the highest probability to the respective emoticon. However, at the moment, we do not consider neutral emoticons. This will be part of future research. 

Comparable to the feelings raised when seeing emoticons, words also have some kind of emotional shade. To identify the emotional shade of the words, we used two databases \footnote{Note that both databases are based on the German language as we performed our application in industry (see Section~\ref{sec:eval}) in a German-speaking company.} summarizing words and their emotionality \cite{remus2010sentiws,waltinger2010sentiment}. The database provided by Waltinger \cite{waltinger2010sentiment} assigns each word to one of the classes \textit{positive (+1), neutral (0)}, or \textit{negative (-1)}, whereas Remus et al.'s \cite{remus2010sentiws} database assigns a value ranging from \textit{-1 (negative)} to \textit{1 (positive)} to each word. We use both databases to increase the number of words and to aggregate the results. Examples from the first database can be found in Table~\ref{tab:databases}. If possible (i.e., if the word is also contained in the second database), we present the concrete value in brackets. 

\begin{table}[t]
	\caption{Examples for the emotional shades \cite{waltinger2010sentiment}. The number in brackets, if available, represent the score \cite{remus2010sentiws}.}	
	\label{tab:databases}
	\centering
	\begin{tabular}{ccc}
		\hline
		\textbf{positive} &  \textbf{neutral} &\textbf{negative} \\ 
		\hline 
		agree (0.0040)  & objectively & arbitrary (-0.3481) \\ 
		convinced (0.2381) & fully & confused \\ 
		innovation (0.0040) & thought &deficient (-0.4535)\\ 
		non-violence &  argumentation &  autocrat \\ 
		\hline 
	\end{tabular}
\end{table}

In addition, we use the \textit{CountVectorizer} and the \textit{TF-IDF Vectorizer} to analyze the relevance of words. For example, the \textit{TF-IDF Vectorizer} calculates the term frequency of the word in the message (TF) and the inverse document frequency (IDF) considering all messages. This helps to detect relevant words which only appear a few times in the text. 

We also take the \textit{formality} of the language into account. However, the detection of formality is language-specific (e.g., in Spanish and German (and other languages) there are specific forms to address someone formally).

\subsection{Step 3: Training the Classifier}
In the next step, we train the classifier to achieve good classification results. We use machine learning techniques to train the classifier, including \textit{Random Forests,} a \textit{Support Vector Machine}, and \textit{Naive Bayes}. To increase the accuracy of our model, we combine the results of the three methods using a \textit{Voting Classifier} choosing the class for a sentence that was forecasted by the majority of the three approaches.

\begin{figure}[htpb]
	\centering
	\includegraphics[width=0.7\linewidth]{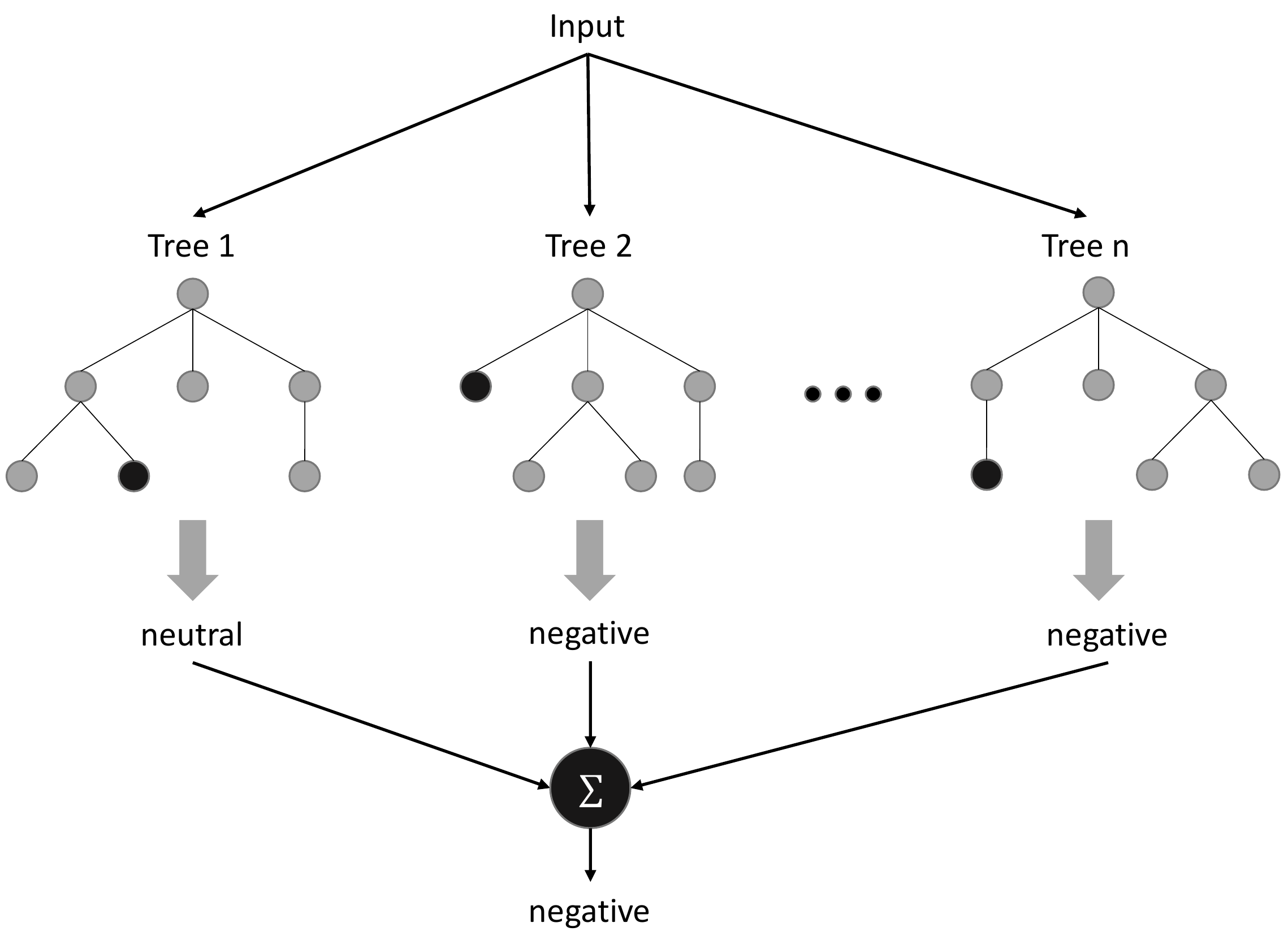}
	\caption{Schematic Visualization of the Random Forest}
	\label{fig:random-forest}
\end{figure}

A \textit{Random Forest} starts with randomly creating decision trees. Each of the trees classifies the sentence. In the end, the algorithm chooses the class that is most often chosen by the trees. An schematic visualization is presented in Fig.~\ref{fig:random-forest}.

\begin{figure}
	\centering
	\includegraphics[width=0.7\linewidth]{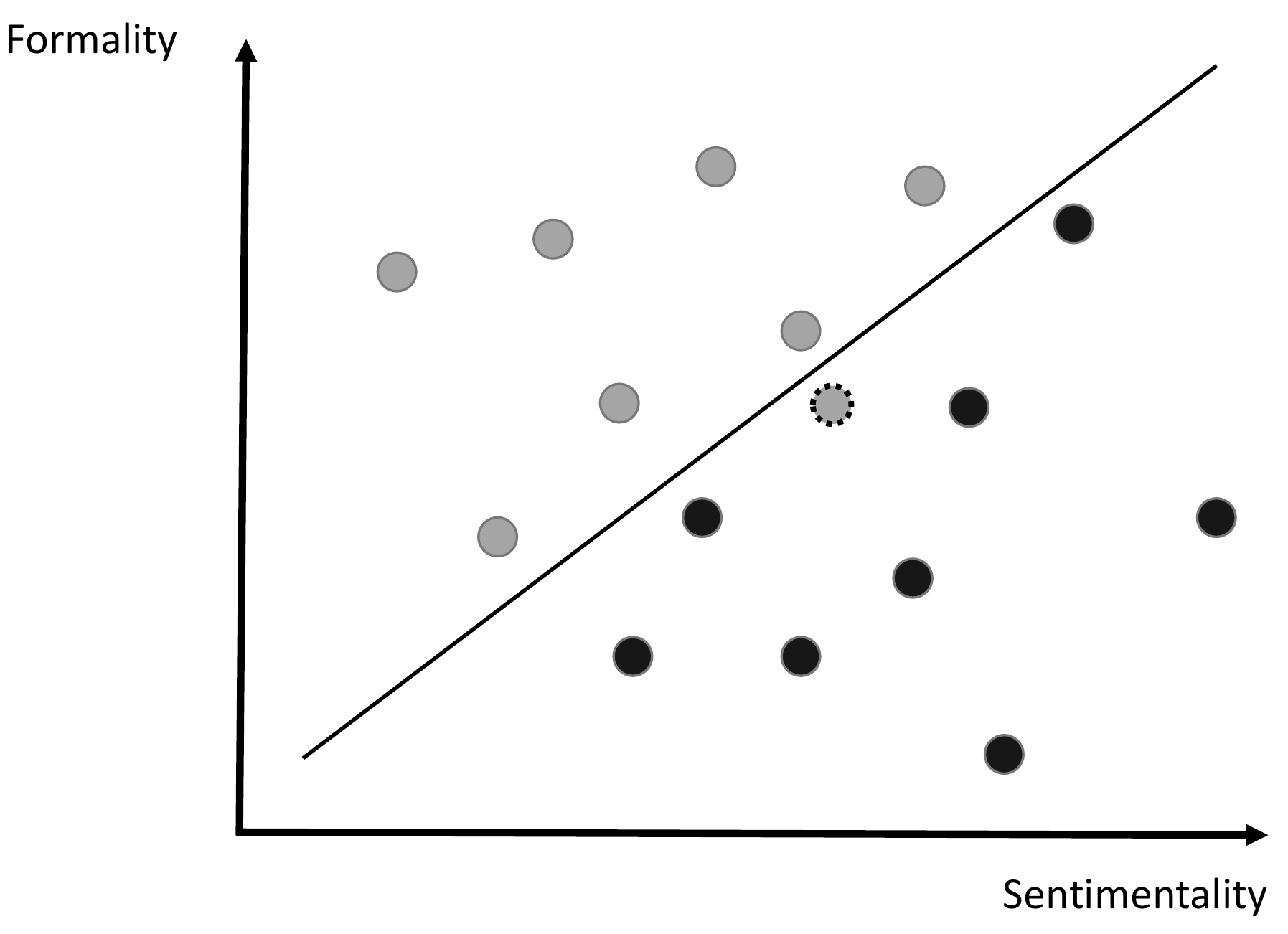}
	\caption{Example for a two-dimensional classification by a Support Vector Machine}
	\label{fig:svm}
\end{figure}

A \textit{Support Vector Machine} separates the multi-dimensional feature space.  Consider the 2-dimensional example in Fig.~\ref{fig:svm}. The gray dots represent data points, i.e., messages, labeled as \textit{neutral}, and the black dots represent data points classified as \textit{negative}. The Support Vector Machine separates this two-dimensional space using a linear function. Therefore, the gray dot circled with black dots would be classified as negative even if the sentence is neutral. 

The \textit{Naive Bayes Classifier} uses probability functions indicating whether a data point in the feature space belongs to a specific class or not. As the density functions of the probability are unknown, this procedure allows only for an approximation. 

As all of the three approaches have their strengths and weaknesses, we decide to combine their results using a \textit{voting classifier}. This classifier identifies the class which was chosen by the majority of the approaches. This way, the approach is more robust against outliers and we can reduce the influence of noise in the data. 

The training of the classifier can be done by dividing the data set into training and test data. The training data is then used to derive heuristics indicating that a specific type of sentence belongs to a specific class (i.e., to train the classifier) and the test data is used to check the accuracy of the classifier. 

To optimize the forecast, we use an evolutionary algorithm. This algorithm uses a fitness function representing the goodness of the current solution, i.e., the classifier. In our case, the goodness is defined via the average accuracy of the classification. We repeat the learning process 20 times with different test and training sets (in a ratio of 10:90). This allows cross-validation of the classifier. 

\subsection{Step 4: Interpretation of the results}\label{sec:interpretation-results}
The interpretation of the results is mainly part of future work. However, to summarize and visualize the emotionality of messages over time, we map the three classes \textit{positive, neutral}, and \textit{negative} to the integer values \textit{+1, 0,} and \textit{$-$1}. This allows to calculate mean values, for example of all messages sent on a specific day, and analyses over time.

\section{Application in Industry}\label{sec:eval}

\begin{figure}
	\centering
	\includegraphics[width=1\linewidth]{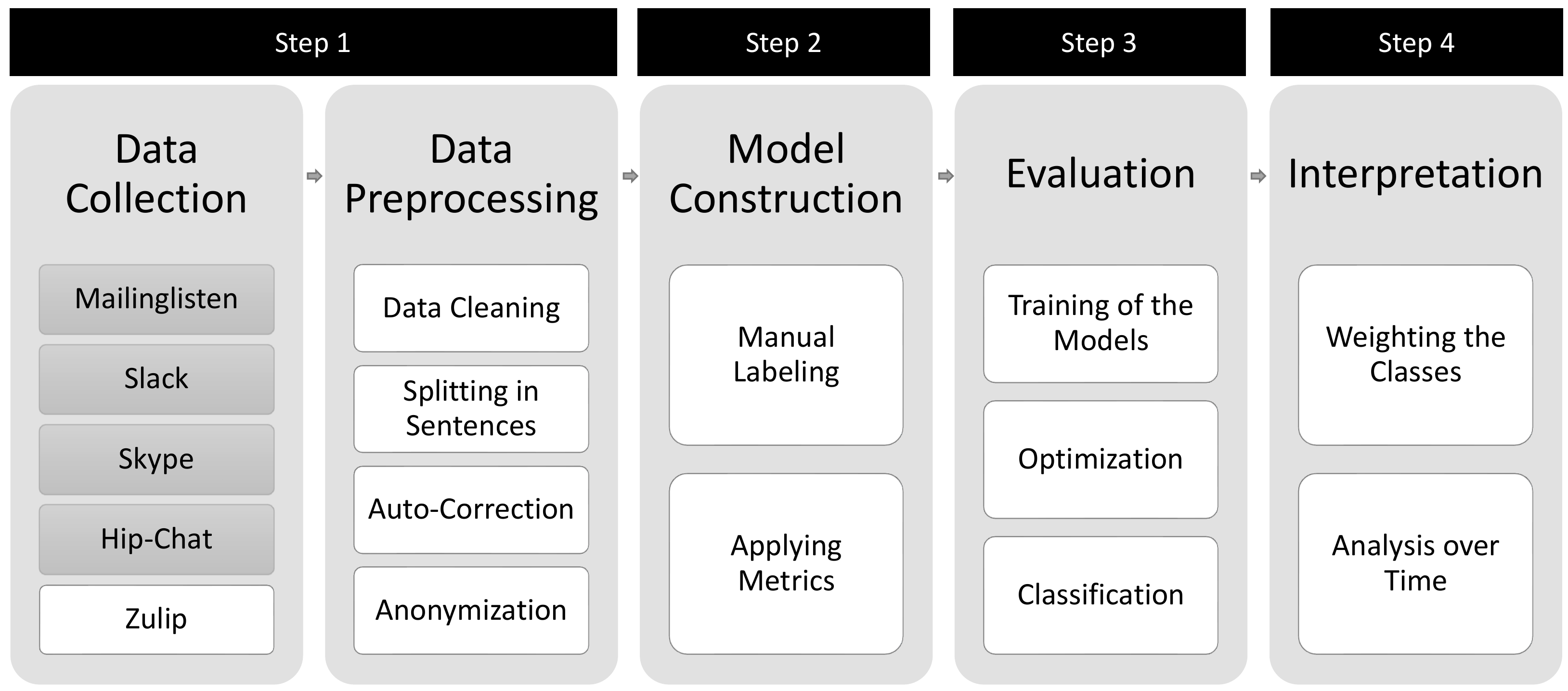}
	\caption{Overview of the application in industry}
	\label{fig:ubersicht}
\end{figure}

To show the applicability of our approach, we apply it to an industrial software project. Figure~\ref{fig:ubersicht} shows the four steps of our general research process with its concrete instantiation in the case study which required a step for the data preprocessing. In the following, we present our proceeding when applying the approach in detail. 

\subsection{Data Collection}
We evaluated our approach in industry. However, due to privacy concerns, we cannot provide profound information on the company. We call the company \textit{ZETA}. \textit{ZETA} is specialized on software and consulting. There is one in-house software development team called \textit{Team MY}. Before running the case study in the company, each team member of \textit{Team MY} received a transparency letter describing the overall concept of our approach as well as a detailed description of the data analysis of communication data of the whole team. After a reflection period of two weeks to communicate concerns, ideas, or disagreement, we were allowed to start the case study since none of the team members disagreed.

In total, about 80 developers worked on the software project we use to prove the applicability of our approach in industry. These developers work distributed in one country. The main communication tool in the software project is \textit{Zulip} which is a chat tool for teams\footnote{More information on Zulip can be found at \url{https://zulipchat.com/}.}. The users subscribe to so-called \textit{streams}. Each stream has a \textit{topic} that organizes conversations in the stream. This helps to cluster messages also after hours of silence on a respective topic. The user can decide on his access to the streams. This reduces the risk of information overload with irrelevant information.  Besides Zulip, the team also uses (if possible) face-to-face communication and meetings, as well as phone calls, e-mails, and a chat tool. However, Zulip is the official tool that should be used for communication.

The company granted us access to five streams of \textit{Team MY} which were exported using the REST-API for further processing. We exported all messages that have been written between Feb 11, 2019 and Jul, 24 2019, resulting in 1947 messages consisting of 7070 sentences. In total, 65 developers actively participated in the communication in at least one of the five streams. Note that we consider a person to actively participate if and only if she wrote at least one message. 

\subsection{Data Processing}
To apply our research process, it was necessary to preprocess the data. The data processing consisted of several steps, starting with the data preparation. This includes (1) data cleaning to remove irrelevant data such as source code which cannot be processed, (2) cutting the messages into sentences, (3) the auto-correction, and (4) anonymization of the messages. 

\begin{itemize}
	\item[(1)] The data export contains several messages respectively strings that cannot be processed. This includes symbols for text highlighting such as asterisks (for bold text) or low lines (for italic text) as well as links and source code. These parts have to be removed. This step was, in our case, done manually. However, we plan to automatize this step in future research.
	\item[(2)] To increase the level of detail of the analysis, the unit of analysis is a single sentence. Therefore, messages which consist of more than one sentence have to be split into single sentences. As start and end of a sentence cannot always be derived by the punctuation, we use the Python package \textit{spacy}. This package identifies sentences based on the structure of the language (and not solely on the punctuation). However, as \textit{spacy} does not have an accuracy of 100\%, we manually checked the results. 
	\item[(3)] Most metrics require the correct spelling of words. Otherwise, words may be identified as a different word mitigating the correctness of the classification. Therefore, we used the Python package \textit{pyspellchecker} providing corrections for misspelled words. 
	\item[(4)] During the anonymization phase, we manually replaced all names and addresses (including links) by pseudonyms. This step was done by one author of this paper. Due to privacy concerns, only one researcher was allowed to process the raw data. Limitations caused by this fact will be discussed in Section~\ref{sec:threats}.  	 
\end{itemize}

\subsection{Training the Model}
In the next step, we trained the model to evaluate the accuracy of the classifier. The training consisted of two steps: (1) Labeling of existing sentences, and (2)~training the classifier.

\begin{itemize}
	\item[(1)] We manually labeled the data by assigning one of the three classes \textit{positive, neutral,} and \textit{negative} to each of the sentences. Examples for words belonging to the respective class are presented in Table~\ref{tab:evalutaion-labelklassen}. The division in three classes is rather coarse-grained. Future research will focus on increasing the granularity of the results. Due to the privacy concerns of the company, this step was done by one researcher. Limitations caused by this fact will be discussed in Section~\ref{sec:threats}. 
	\item[(2)] In the training phase, the metrics presented in Section~\ref{sec:metrics} are calculated for each of the sentences. We calculated in total 5378 different metrics, mostly considering language-specific characteristics\footnote{As the number of metrics is quite high, future research should focus on the selection of appropriate metrics.}. We chose this huge number of metrics as each metric which is not correlated to other metrics can increase the accuracy of the classifier. To analyze the correlation between the metrics, we calculated the correlation matrix (based on Pearson's $r$) for this specific data set.  
\end{itemize}

\begin{table}[!ht]
	\caption{Exemplary emotions or types of words for each class}	
	\label{tab:evalutaion-labelklassen}
	\centering
	\begin{tabular}{ccc}
		\hline
		positive &  neutral & negative \\
		\hline
	love	&facts  & fear \\
	happiness	& ambivalence &  hate\\
	euphoria	& interest & anger  \\
	surprise	& indifference &  trouble \\
	sympathy	& apathy &  regret\\
		\hline
	\end{tabular}
\end{table}

\subsection{Evaluating the Model}
We evaluate the model by applying the classifier to some exemplary sentences of our data set. We chose a ratio of 10\% for the test data set and use the remaining 90\% of the data set to train the model. The selection of the concrete test data was random. The learning process of the model is based on an evolutionary algorithm which identifies the best fitting model. In our case, we achieved the best model after 80 steps. The accuracy of the model is defined by the confusion matrix summarizing the results of the classification (comparison between forecast and as-is) as well as different key indicators such as precision, recall, and F1-score. 

\subsection{Interpretation of the results}
To visualize the results, we present the development of the average emotional score of the messages over time. We assign values to each of the classes, namely \textit{positive = +1}, \textit{neutral = 0}, and \textit{negative = -1}. This way, it is possible to calculate the average score per day, which we call the \textit{emotionality score}. Observing this score over time can help to detect stressful phases. However, the interpretation of the emotionality score remains future work.

\section{Results}\label{sec:results}
We performed the steps described in Section~\ref{sec:eval}. The manual data labeling resulted in 845 sentences classified as \textit{positive}, 1856 sentences classified as \textit{neutral}, and 1077 sentences classified as \textit{negative}. This imbalance of data points can influence the classifier's accuracy as there are more negative sentences to train the classifier than positive. However, it is unlikely to find completely balanced data in industry. Future research should, hence, focus on training the model with huge data sets containing balanced data. 

\begin{figure}
	\centering
	\includegraphics[width=1\linewidth]{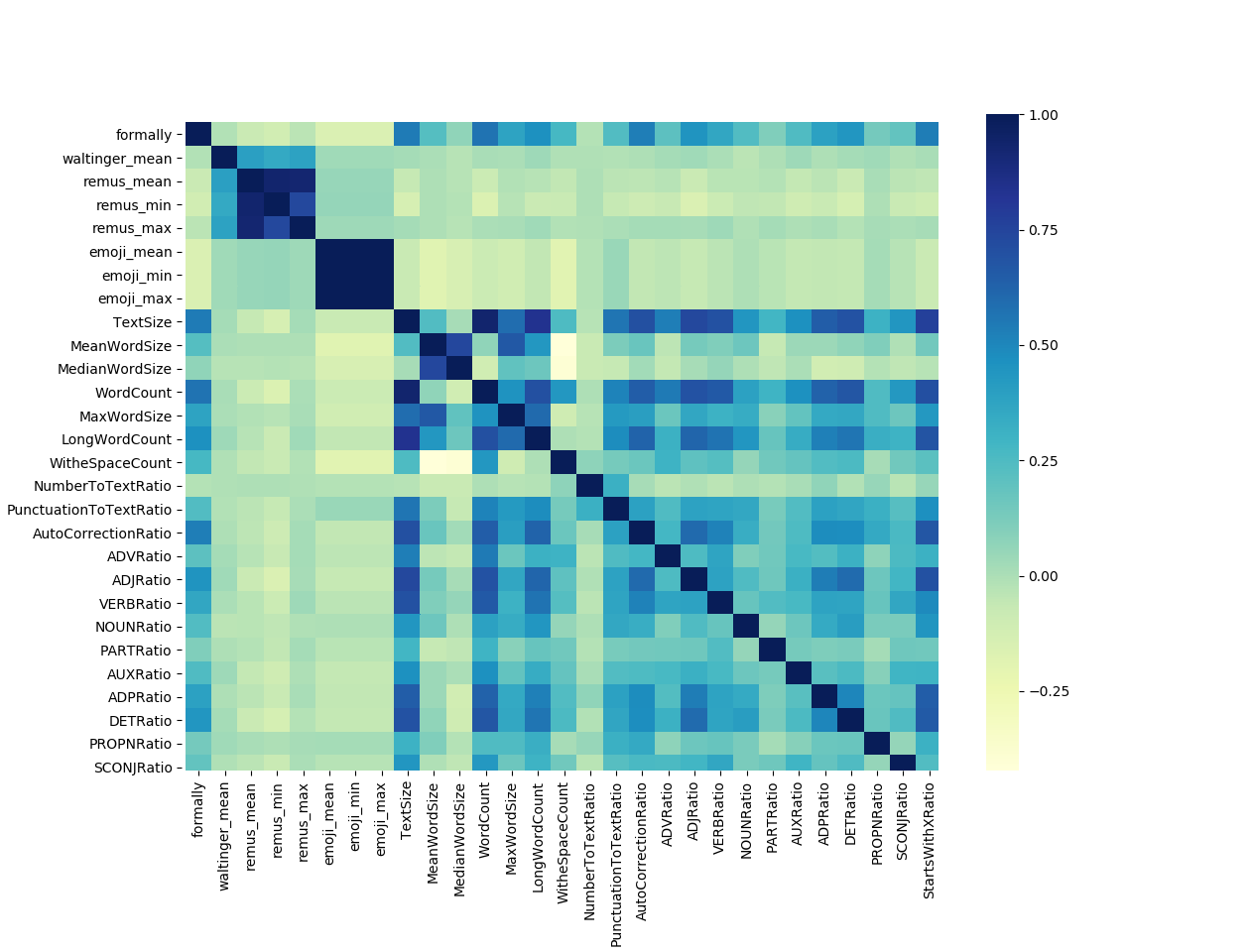}
	\caption{Correlation matrix including 28 metrics}
	\label{fig:correlation-matrix}
\end{figure}

In a next step, we calculated the correlation matrix of the metrics presented in Fig.~\ref{fig:correlation-matrix}. Most of the metrics are only partially or not at all correlated. However, the metrics for the emotions (emoji\_mean, emoji\_min, and emoji\_max) are highly correlated. In addition, we find a weak correlation between the need for corrections (AutoCorrectionRatio) and the formality of the communication (formality). 

\begin{table}[!thpb]
	\caption{Confusion matrix of the classifier for the test set}	
	\label{tab:evalutaion-konfusionmatrix}
	\centering
	\begin{tabular}{c|c|c|c|}
		& \tabrotate{\ \textbf{positive}} & \tabrotate{\ \textbf{neutral}} & \tabrotate{\ \textbf{negative}}\\		
		\hline 
		\textbf{positive} & 36 & 46 & 7\\
		\hline 
		\textbf{neutral} & 10 & 173 & 11\\ 
		\hline 
		\textbf{negative} & 2 & 64 & 29\\ 
		\hline 
	\end{tabular} 
\end{table}

As described above, we trained the model using the labeled data set. The classifier assigned one of the classes \textit{positive, neutral,} and \textit{negative} to the sentence under consideration. The evolutionary algorithm used to train the model achieved, in his best generation, an average accuracy of 58.5\%. We chose the best model from this generation which achieved an average accuracy of 62.97\%. To calculate the accuracy of our model, we applied an amount of randomly chosen 10\% of the data set as test data to the model trained by the remaining 90\% of the data set as training data. The results of the classification are presented in Table~\ref{tab:evalutaion-konfusionmatrix} as a confusion matrix. We see only 9 out of 378 confused classifications of positive and negative, i.e., it was most difficult to distinguish between positive and neutral (56 out of 378 confusions) respectively between negative and neutral (75 out of 378). Based on these numbers, it is possible to calculate other key indicators summarized in Table~\ref{tab:evalutaion-klassifizierungsreport}. In total, the class \textit{positive} has the highest precision, whereas the F1-Score is best for the neutral class. Examples for a few classifications are presented in Table~\ref{tab:evalutaion-prediction-sampels} presenting the manual classification by a person and the predicted classification by the model.

\begin{table}
	\caption{Classification report for the test set}	
	\label{tab:evalutaion-klassifizierungsreport}
	\centering
	\begin{tabular}{ccccc}
		\hline 
		Class & Precision & Recall & F1-Score & Frequency \\
		\hline 
		positive & 0.75 & 0.40 & 0.53 & 89 \\  
		neutral & 0.61 & 0.89 & 0.73 & 194 \\
		negative & 0.62 & 0.31 & 0.41 & 95 \\ 
		\hline 
	\end{tabular} 
\end{table}

\begin{table}[H]
	\caption{Exemplary sentences and predictions from a human rater and the trained classifier}	
	\label{tab:evalutaion-prediction-sampels}
	\centering
	\begin{tabular}{clcc}
		\hline 
		Id & Content & Human & Classifier \\ 
		\hline 
		1 & Yes, this was my mistake. & negative &neutral\\ 
		2 & Welcome in [[countryname]] :-) good decision & positive & positive\\ 
		3 & If this was not your mistake, it has to be fixed as follows: & neutral & neutral\\ 
		4 & Had understood you differently this morning. & negative & neutral\\ 
		5 & Then we agree. & positive & neutral\\ 
		6 & Well done! & positive & positive\\ 
		\hline 
	\end{tabular} 
\end{table}

\begin{figure}[!t]
	\centering
	\includegraphics[width=0.9\linewidth]{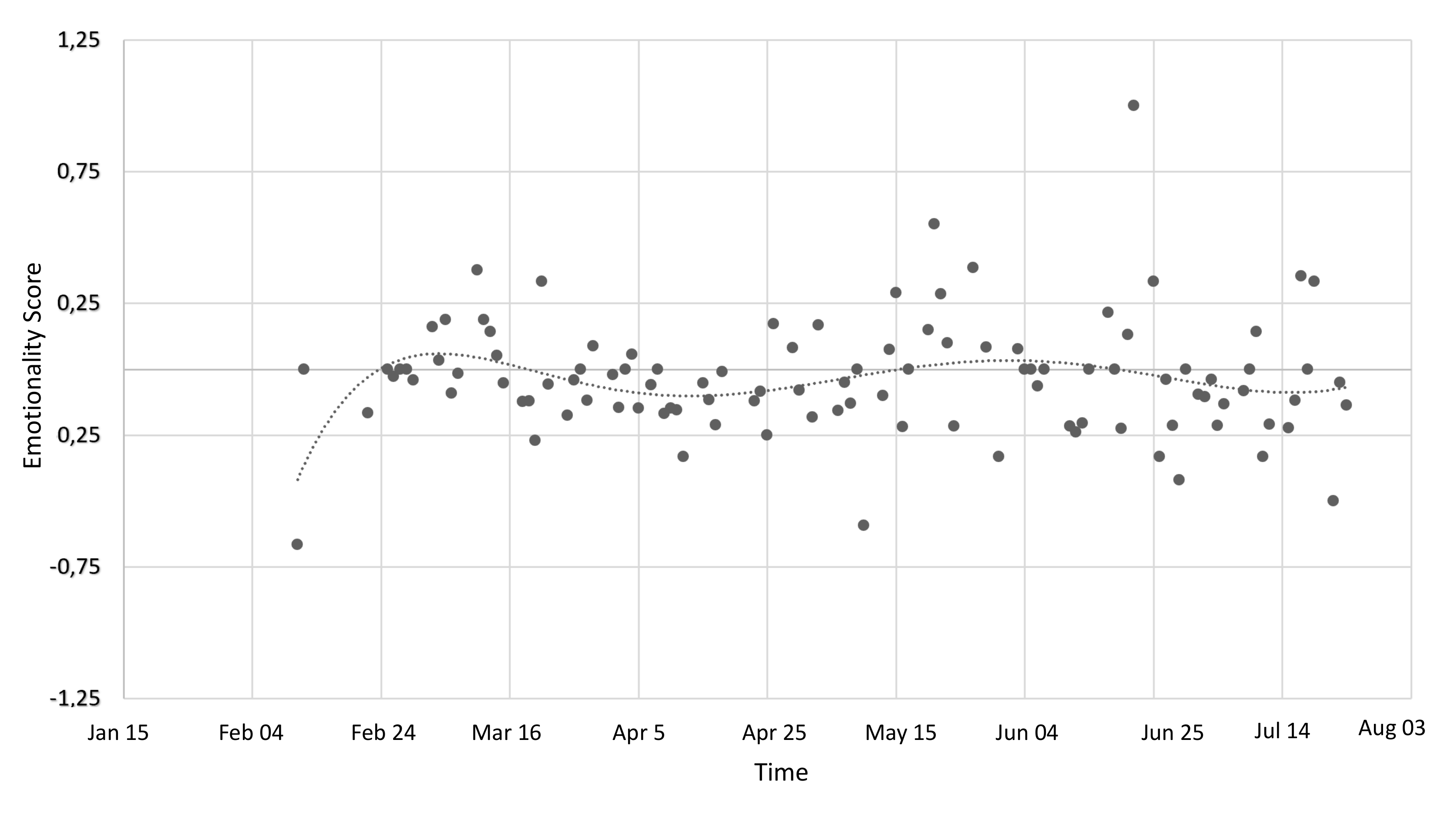}
	\caption{Emotional niveau using the manually labeled data. The dotted curve presents the trendline.}
	\label{fig:emotion-over-time}
\end{figure}

\begin{figure}[!h]
	\centering
	\includegraphics[width=0.9\linewidth]{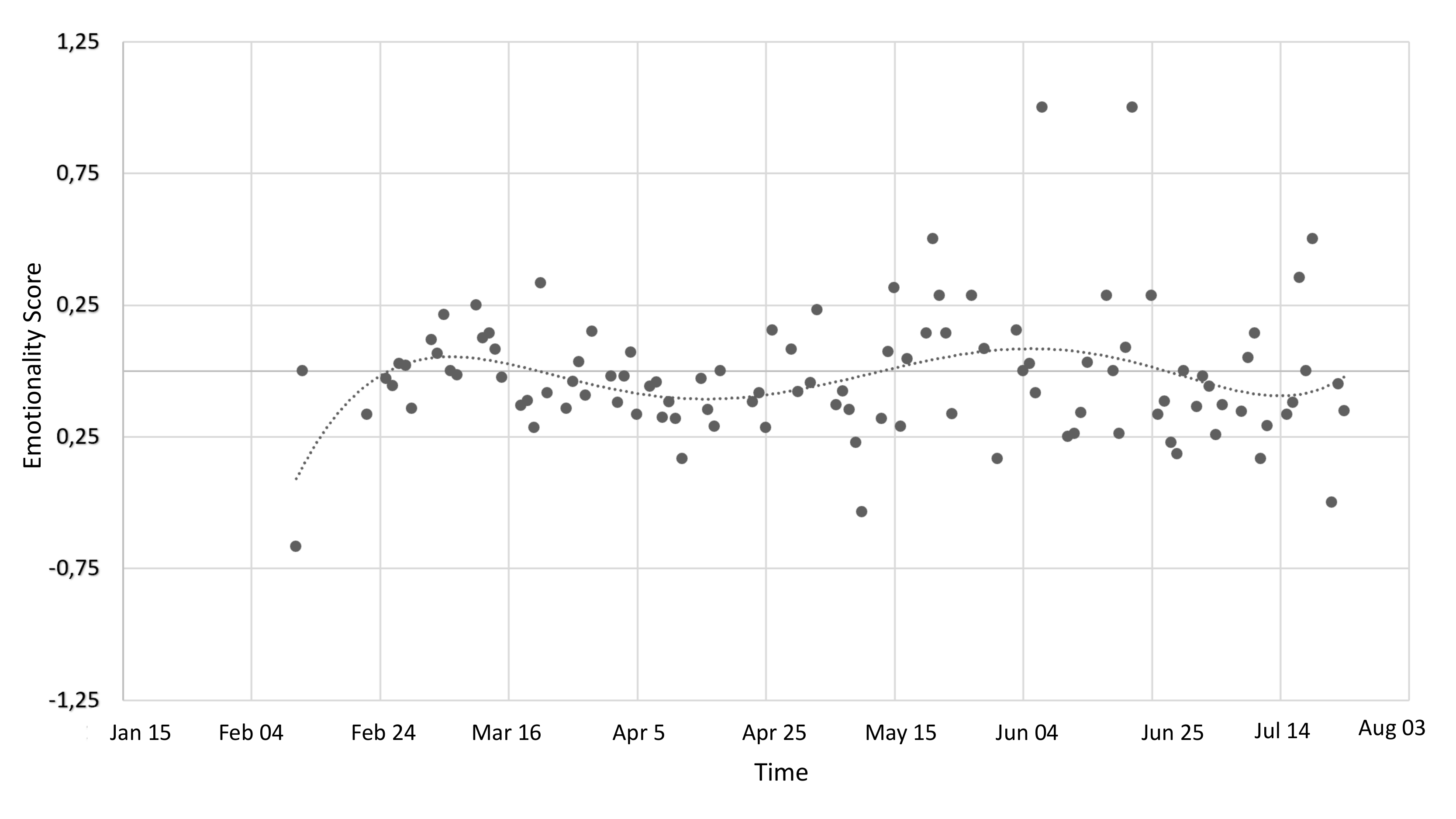}
	\caption{Emotional niveau using the forecasted data. The dotted curve presents the trendline.}
	\label{fig:emotion-over-time-fc}
\end{figure}

At this point, we have evaluated the classifier leading to an accuracy of 62.97\%. In the last step, we want to outline how the classification (either from a person or from the classifier) can be used to analyze emotions in the team. Presenting the shade of the emotions on a scale from -1 to 1 as described in Section~\ref{sec:interpretation-results} leads to the curve presented in Fig.~\ref{fig:emotion-over-time}. Note that we use the manually labeled data. However, this step can also be done using the results of the classifier (see Fig.~\ref{fig:emotion-over-time-fc}). To derive the results in Fig.~\ref{fig:emotion-over-time}, we first calculate the average emotional score of the labeled data on a daily level. As evident from Fig~\ref{fig:emotion-over-time}, the mood is on average rather neutral. However, it ranges from positive to negative and vice versa. When presenting the results to the company, we figured out that the negative phases coincide with stressful phases in the project, e.g., due to deadlines. Comparing the trendline in Fig.~\ref{fig:emotion-over-time} with the trendline in Fig.~\ref{fig:emotion-over-time-fc} based on the forecasted data shows very similar tendencies. 

\section{Discussion}\label{sec:discussion}\label{sec:threats}
In the application in industry, we analyzed text-based communication consisting of 1947 messages in a chat tool used by a development team in industry. We used several metrics ranging from the emotional shade of single words over the length of a sentence to the formality of the language. We combined three machine learning algorithms to classify each of the messages as \textit{positive, neutral}, or \textit{negative}. 

Our approach achieved an accuracy of 62.97\% which allows for improvement. However, the trained classifier was almost as good as a human rater who achieved an accuracy of 66\% when coding 200 sentences twice with a temporal distance of one month. 

We used the classification to aggregate the team mood on a daily level by calculating the average emotionality of all messages sent on a specific day. When interpreting the course of emotionality over time, we see an average neutral mood, with some tendencies towards positive as well as negative mood. However, one would expect to find such a rather balanced course in a professional work environment. Further possibilities to interpret the results will be subject to future research. 

\subsection{Threats to Validity}
Our application in industry is subject to some threats to validity which we discuss now. 

The \textit{conclusion validity} is threatened by the choice of the data source. For example, communication in e-mails will differ from communication in group chats. We are aware that adjustments to the procedure are required, as the data collection and extraction strongly depends on the data source. To provide a holistic picture of the emotionality of the text-based communication in the team, we consider a huge set of metrics. To allow good learning of the model, we ensure that these metrics are only partially correlated and do not measure all the same characteristics of the communication. 

Due to privacy concerns and legal restrictions from the company, only one author was allowed to handle the data. This threatens the \textit{internal validity} since the data processing strongly depends on the subjective perception of one person. By following a structured proceeding, for example for the anonymization of the data, we tried to mitigate this threat. However, there was no possibility to review the labeling process. Therefore, the person who labeled the data labeled a randomly chosen set of 200 sentences a second time with a distance of one month. In 2 of 3 cases, the labels coincided, leading to an accuracy of 66\%. This raises several questions and possibilities for future research which will be discussed in Section~\ref{sec:fw}. However, the manual reference classification of the data influences the results as this defines whether the classification by the tool is correct or not. 

The \textit{construct validity} may be threatened due to the mono-operation and the mono-method bias. As we are aware of this fact, we do not draw any conclusions based on our results. We only see them as promising support to continue the development of the approach. Further studies in different settings are required to mitigate this threat. 

The \textit{external validity} of our results is very limited. The concrete results of our analysis are correct for the team under consideration, and only reflect the time frame of our analysis. However, we do not want to draw conclusions or to generalize our results for other teams. The main objective of the application to an industrial project is to prove the applicability of our approach. If this is the case, we only conclude that our approach is applicable, but the results obtained from the study cannot be transferred to any other team. How to draw conclusions based on the results will be subject to future work (see Section~\ref{sec:fw}).

\subsection{Answering the Research Question}
With the approach presented in this paper and its application in industry, we were able to show that the analysis of text-based communication is possible using a combination of natural language processing, sentiment analysis, and machine learning. By aggregating the emotionality of the sentences, e.g., daily, it is possible to derive information on the general mood in the team as well as the development over time. However, there are some open questions and there is potential for improvement to be addressed in future work.  

\subsection{Future Research}\label{sec:fw}
Even if the results of our approach underline its potential, it might (and should) be improved to address some issues that could not have been considered in the paper at hand. These issues will be addressed in future research.

(1) One problem is the rather low accuracy of the algorithm. However, given the interrater agreement of a human rater of 66\% when labeling the data twice with a temporal distance of one month, the classifier is almost as good as the human rater. This result is near to perfect as a trained model cannot perform better than the labeling allows. Consequently, to improve the accuracy of our algorithm, we first have to find possibilities to identify clear characteristics of the messages in the respective class. This finding goes along with the findings of other authors \cite{jongeling2015choosing}.

(2) The context of the single sentence is currently not considered. This should be improved in future research. Depending on the context of previous messages, an answer can be differently interpreted. Therefore, the emotionality of the sentences before also needs to be considered for the forecast and probably can also improve the classification. 

(3) At the moment, we do not provide any guidance for the interpretation of the results. To gain insights based on the results, we calculate average emotionality scores and present them on a daily level. This can help to detect stressful phases or phases where the team needs support, for example by an external coach due to unsolved team-internal conflicts. However, as part of future research, it is required to apply our approach to huge data sets and investigate correlations to other factors (e.g., deadlines, conflicts, high workloads, and context factors). 

(4) The usefulness of the tool for development teams is not yet proven. This is mainly because we do not provide guidance to interpret the results. As soon as we finished the research described in (3), we can evaluate the usefulness of the insights by presenting them to the team. 

\section{Conclusion}\label{sec:conclusion}
As text-based communication is widely distributed in software development, but often not adequate, we strive towards an automated analysis of text-based communication in different channels. Our approach analyzes interpersonal behavior in text-based communication concerning the emotional shade of the communication, considering the conversational tone, the familiarity of sender and receiver, and the appropriateness of the used language. We prove the applicability of our approach in an industrial case study, where we got access to 1947 messages in Zulip, consisting of 7070 sentences. The results of our application in industry show that it is possible to correctly classify statements with an average accuracy of 62.97\% which is as good as the rating of a human classifier. In this paper, we only classified the statements as \textit{positive}, \textit{neutral}, or \textit{negative}. In future work, we want to make the results more fine-grained and also consider other information contained in the data, such as emoticons. In addition, we want to support the interpretation of the results, which is not yet part of the approach.

\bibliographystyle{splncs04}
\bibliography{references}

\end{document}